# Moderating effects of gender and family responsibilities on the relations between work-family policies and job performance

José Aurelio Medina-Garrido, José María Biedma-Ferrer and Antonio Rafael Ramos-Rodríguez

INDESS (Instituto Universitario para el Desarrollo Social Sostenible), Universidad de Cádiz, Spain

This is the preprint version accepted for publication in the " International Journal of Human Resource Management". The final published version can be found at:

https://doi.org/10.1080/09585192.2018.1505762

We acknowledge that Routledge Journals, Taylor & Francis LTD holds the copyright of the final version of this work. Please, cite this paper in this way:

Medina-Garrido JA, Biedma-Ferrer JM, Ramos-Rodríguez AR (2021) Moderating effects of gender and family responsibilities on the relations between work–family policies and job performance. International Journal of Human Resource Management, 32:1006–1037.

## Abstract

This study analyzes the impact of work-family policies (WFP) on job performance, and the possible moderating role of gender and family responsibilities. Hypothesis testing was performed using a structural equation model based on a PLS-SEM approach applied to a sample of 1,511 employees of the Spanish banking sector. The results show that neither the existence nor the accessibility of the WFP has a direct, positive impact on performance, unlike what we expected, but both have an indirect effect via the well-being generated by these policies. We also find that neither gender nor family responsibilities have a significant moderating role on these relations, contrary to what we initially expected.

## Keywords

Work-family conflict, work-family policies, work-family balance, family-friendly policy, job performance, employee well-being, gender, family responsibilities, human resource management, HRM.

**Introduction**

Having policies in place to help workers balance work and family—work-family policies (henceforth, WFP)—is an intangible capital within corporate social responsibility because it generates an improvement in the firm's external social image. But WFP could also have positive repercussions for workers' performance (Medina-Garrido, Biedma-Ferrer, & Ramos-Rodríguez, 2017; Biedma-Ferrer & Medina-Garrido, 2014; Mills, Matthews, Henning, & Woo, 2014). Thus analyzing the relation between WFP and job performance is of particular interest to organizations and hence researchers in the field of human resource management.

Work-family conflict (henceforth, WFC) arises when the responsibilities associated with work interfere with the employee's activities in their family (Lin, 2013). According to Pareek's (2002) types of role stress, WFC can be classified as an inter role distance. This type of role stress arises as a result of the various roles that an individual occupies in the organization and non-organization setting (Pareek, 2002). WFC is a stressor that could have a negative impact on job performance and hence ultimately on organizational performance (Li, Bagger, & Cropanzano, 2016; Greenidge & Coyne, 2014; Beauregard & Henry, 2009; Yasbek, 2004).

An adequate management of the human resources would therefore seem to require a thorough analysis and appropriate management of WFC. Thus a number of organizations have shown great interest in this conflict, and have adopted WFP to try to reduce it. These policies include a wide variety of practices that help workers to balance the demands of their work and the obligations linked to their family life (Lobel & Kossek, 1996).

The literature finds that WFC leads to a lower organizational performance (Sánchez-Vidal et al., 2011). When workers perceive that the situation is unfair or threatens their well-being, WFC generates stress and negative emotions that are predictors of counter-productive behaviors such as reduction of effort and deliberately doing tasks incorrectly (Matta et al., 2014).

In turn, WFP are considered to make a positive contribution to organizational performance (Barnett, 1998; Clark, 2000; Hill et al., 2001; Anderson et al., 2002). The employees who are more satisfied with WFP will have a higher level of job satisfaction, commitment, and performance (Greenhaus and Powell 2006; Carlson, Grzywacz, & Zivnuska, 2009; de Sivatte & Guadamillas, 2013; Ko et al., 2013).

In this line, our objective with this work is to analyze the impact of WFP on job performance. For this purpose, we propose an unprecedented model explaining the effects of WFP on employees' well-being in the workplace and hence ultimately on their job performance.

Furthermore, for the WFP to actually generate benefits the workers need to know that they exist (Yeandle et al., 2002; Haar & Spell, 2004; Mumford, 2005), but they also need to perceive that the policies are accessible to them without negative consequences for themselves (Gray & Tudball, 2002; Whitehouse & Zeitlin, 1999; Bond, 2004). Because the mere existence of WFP is insufficient (Yeandle et al., 2002; Budd & Mumford, 2005), we will differentiate in our analysis between whether such policies exist and are known about by the employees, on the one hand, and whether they are accessible in practice to the workers, on the other. In this sense, the study makes a significant theoretical and empirical contribution to the human resource field by analyzing separately the impact of WFP existence and accessibility on well-being and performance. This empirical analysis constitutes a significant contribution to the literature, where a number of works distinguish between WFP existence and accessibility (Gray & Tudball, 2002; Whitehouse & Zeitlin, 1999; Bond, 2004; Yeandle et al., 2002; Budd & Mumford, 2005), but there are no empirical studies analyzing their impact on employee well-being and performance.

Finally, the literature stresses that the WFP-performance relationship may differ for female workers (Harrington, 2007; Harris, 2004; Kirkwood & Tootell, 2008) and workers with family responsibilities (Premeaux et al., 2007). So we also analyze the possible moderating role of these two variables in the above relations.

This work is organized as follows. The next section establishes the theoretical foundations on which we develop our hypotheses. The third section describes our empirical analysis and the methodology used and analyzes the results we obtain. The fourth section discusses the results. Finally, the final section offers the conclusions of this study and suggests the limitations of the work and possible future lines of research.

**Theoretical foundations and hypotheses**

The literature examines the impact of WFC and WFP on performance (Drago et al., 2001; Estes et al., 2007; Glass & Estes, 1997; Swody & Powell, 2007; Barnett, 1998). A number of studies emphasize that WFC has a negative impact on employee performance (Sánchez-Vidal et al., 2011; Matta et al., 2014). According to the theory on appraisals and coping (Lazarus and Folkman, 1984), the cognitive appraisal of the WFC could lead to responses such as reducing job satisfaction, increasing psychological strain or distress, increasing intent to quit, reducing effort or deliberately doing tasks incorrectly.

On the contrary, literature considers that WFP alleviate the tension between work and family, increasing employees' productivity (Estes et al., 2007; Swody & Powell, 2007), and making a positive contribution to organizational performance (Barnett, 1998; Ko et al., 2013).

In many cases, literature assumes that the WFC and WFP impacts on employee performance are direct, but some studies consider the possible intervention of mediating variables that—influenced by WFC or the WFP in place—ultimately affect job performance. The most widely studied of these mediating variables are stress, job satisfaction, motivation, and commitment to the organization (Ravangard et al., 2015; Wang et al., 2013; Karatepe, 2013; Hyun & Kim, 2012). These variables can generate both positive and negative affective states that influence the worker's psychological well-being (Higgs & Dulewicz, 2014), so they can be included within the concept of emotion-based well-being (Wright, 2014). And the literature on workers' emotion-based well-being finds that this well-being has a significant effect on job performance (Wright, 2014).

The need to reduce the problems resulting from WFC has encouraged firms to implement WFP. But having these policies in place is not enough: the workers need to know about them (Budd & Mumford, 2005; Yeandle et al., 2002) and to be able to access them without negative repercussions for themselves (Whitehouse & Zeitlin, 1999; Gray & Tudball, 2002; Bond, 2004).

WFP are mainly designed with workers who have family responsibilities in mind (Premeaux et al., 2007). The literature observes that women assume these responsibilities to a greater extent than men, so gender is an important variable in the WFC literature (Blanch & Aluja, 2012; Linehan & Walsh, 2000; Harrington, 2007; Harris, 2004; Kirkwood & Tootell, 2008).

We now describe the theoretical framework that will form the foundations for the hypotheses in our model that we statistically test in our subsequent empirical study.

*WFP: existence and accessibility*

In the WFP literature authors often analyze the impact of the existence of such policies on workers' behaviors and attitudes (Sturman & Walsh, 2014; Vanderpool & Way, 2013). But the mere existence of such policies is a necessary but insufficient condition. For the WFP to actually generate benefits the workers first need to know that they exist (Yeandle et al., 2002; Mumford, 2005). They then need to perceive that the policies are accessible to them, in other words they can use them without negative consequences for themselves (Gray & Tudball, 2002; Whitehouse & Zeitlin, 1999; Bond, 2004).

Some studies examining the accessibility of WFP suggest that workers are not always aware that they can use these policies (Yeandle et al., 2002; Budd and Mumford, 2005). Moreover, the fact that WFP are in place in the firm does not necessarily mean that the workers perceive that they can use them freely (Gray & Tudball, 2002; Whitehouse & Zeitlin, 1999; Bond, 2004). For this, the organization needs to promote a climate that supports the use of the WFP. Workers may be discouraged from using these policies out of fear that they will lose their chances of promotion, that others will think they are not fully committed to the firm, or even that they will lose their jobs (Whitehouse & Zeitlin, 1999).

The hypotheses we propose in this work take into account this twofold dimension of WFP. On the one hand, they consider the existence of these policies and the workers' knowledge of their existence. And on the other, they consider to what extent the WFP are accessible without subsequent reprisals or any other type of negative repercussions.

*Influence of family responsibilities and gender*

By definition, the workers who are most likely to suffer from WFC are those with some type of family responsibility (Premeaux et al., 2007). Kossek et al. (2001) empirically demonstrate the impact of the family responsibilities of looking after children or elderly relatives on WFC, and on well-being and job performance. Caregivers who worked in supportive organizations (e.g. adopting WFP) had higher odds of good work outcomes (Plaisier et al., 2014). Nevertheless, according to de Janasz, Forret, Haack, & Jonsen (2013), by providing a more 'family-friendly' work environment, organizations may foster perceptions of inequity in people without family responsibilities, possibly resulting in lower job satisfaction and other work-related outcomes.

The theory recognizes gender differences in the assumption of family responsibilities and consequently in the probability of experiencing WFC (Harris, 2004; Harrington, 2007; Kirkwood & Tootell, 2008; Baral & Bhargava, 2010; Cloninger, Selvarajan, Singh, & Huang, 2015; Foster & Ren, 2015). Harris (2004) argues that as a result of WFC women are likely to suffer more stress and to show lower job satisfaction than men.

According to Gutek et al. (1991), gender role theory establishes that family and work roles have traditionally been gender specific, where men are work oriented and women family oriented. Thus, a number of studies support the idea that the negative relation between WFC and job satisfaction is stronger among women than among men (Bruck et al., 2002; Kossek & Ozeki, 1998). Some researchers in contrast fail to find empirical evidence for the existence of significant differences in the impact of gender on the existence of WFC (Akintayo, 2010; Hughes & Bozionelos, 2007).

Thus it seems advisable to analyze the possible moderating effects of gender and family responsibilities on the relations between WFP and the worker's behavior (performance, absenteeism or quitting) and psychological state (stress, satisfaction, motivation or commitment). In this respect, our hypotheses take into account the possible moderating role of family responsibilities and gender.

*WFP and performance*

The literature points out that WFC leads to a lower organizational performance (Sánchez-Vidal et al., 2011). Matta et al. (2014) argue that significant work events, such as WFC, in which the worker perceives that the situation is unfair or threatens their economic or social well-being, generate stress and negative emotions and affective states. According to these authors, these are predictors of counter-productive behaviors such as reduction of effort, deliberately doing tasks incorrectly, verbal and physical abuse, and theft (Penney & Spector, 2005; Spector et al., 2006).

The negative emotions generated by a WFC and their negative impact on employee performance may be drawn from Lazarus and Folkman's (1984) theory on appraisals and coping. Cognitive appraisal is the process where an individual evaluates whether a WFC is relevant to his or her wellbeing and, if so, there is an evaluation of the coping options, in other words, of what can be done about this WFC. There are two well-known types of coping: emotion-focused and problem-focused (Lazarus & Folkman, 1984). Thus WFC leads to emotional responses such as reduced job satisfaction (Glaveli, Karassavidou, & Zafiropoulos, 2013), increased psychological strain or distress, and increased intent to quit, among other negative consequences, potentially leading to increased emotional exhaustion (Rodwell et al., 2014). Besides, WFC could also lead to problem-focused responses (options for coping with the situation) such as reducing effort or deliberately doing tasks incorrectly.

In turn, WFP are considered to make a positive contribution to organizational performance (Barnett, 1998). WFP improve job satisfaction and organizational commitment, factors that are closely related to employee productivity and performance

(Greenhaus and Powell 2006). According to Ko et al. (2013) the employees who are more satisfied with WFP will have a higher level of job satisfaction and performance. Hence, when firms alleviate the tension between work and family, worker productivity increases (Estes et al., 2007; Swody & Powell, 2007).

The first hypotheses of this work follow:

H1: The greater the worker's perception that WFP exist in their firm, the greater their job performance.

H1a: The relation between the perception of the existence of WFP and job performance is stronger in women than in men.

H1b: The relation between the perception of the existence of WFP and job performance is stronger in workers with family responsibilities than in workers without family responsibilities.

H2: The greater the worker's perception that the WFP are accessible, the greater their job performance.

H2a: The relation between the perception of the accessibility of the WFP and job performance is stronger in women than in men.

H2b: The relation between the perception of the accessibility of the WFP and job performance is stronger in workers with family responsibilities than in workers without family responsibilities.

***WFP and well-being***

Wright (2014) points out that researchers have been interested in the role of employee well-being, but also indicates that the literature has not yet reached a consensus about its definition, components or how to measure it (see e.g. Knight & Haslam, 2010; Loretto, Platt, & Popham, 2009; Wegge, van Dick, Fisher, West, & Dawson, 2006; van Dick, Ullrich, & Tissington, 2006; Nicholson & Imaizumi, 1993). Adopting Cropanzano and Wright's (2014) classification, the contributions to the literature on this question can be grouped into four categories: (1) objective living conditions (wealth, health, status, education, etc.); (2) eudaimonic well-being (related to having a purpose in life); (3) satisfaction with life; and (4) emotion-based well-being. The literature on WFC often analyzes the impact of this conflict on a number of psychological variables that can be included in the concept of employee well-being. The most common variables examined include job satisfaction, which is assigned to Cropanzano and Wright's (2014) third category, and stress, motivation and commitment, which can be placed in the fourth category, known as emotion-based well-being. The values of these psychological variables can be influenced by WFC and WFP (Medina-Garrido et al., 2017; Biedma-Ferrer & Medina-Garrido, 2014; Beauregard, 2011; Redman, Snape, & Ashurst, 2009). And considering these variables as components of employee well-being, the literature establishes a negative correlation between WFC and employee well-being (Lapierre & Allen, 2006; Fiksenbaum, 2014).

With regard to job stress, many studies argue that WFC causes stress (Baxter & Chesters, 2011; Zhang & Liu, 2011; Anderson et al., 2002) and hence negatively affects the individual's well-being (Martha, 2010). The literature also finds a negative relation between WFP and stress (Litchfield et al., 2004), and a positive relation between lack of WFP and stress (Idrovo, 2006).

On the other hand, research on WFC does not pay sufficient attention to motivation (Zhang & Liu, 2011). But some evidence does exist for a positive relation between WFP and motivation (López-Ibor et al., 2010; Meil et al., 2008).

Job satisfaction—understood as the feelings individuals have about their work (Locke, 1976)—can be considered an important indicator of employee well-being (De Jonge et al., 2000; Page & Vella-Brodrick, 2009). Researchers have found a negative relation between WFC and job satisfaction (Glaveli et al. 2013; Premeaux et al., 2007;

Bagger et al., 2008; Carr et al., 2008; Anderson et al., 2002; Wolfram and Gratton, 2014), and a positive relation between WFP and job satisfaction (Allen, 2001; Batt & Valcour, 2003; Aryee et al., 2005; Kinnunen et al., 2005; Balmforth & Gardner, 2006; Frye & Breaugh, 2004; Behan & Drobnic, 2010).

Finally, the existence of commitment to the organization can also be considered an indicator of the level of employee well-being. Organizational commitment can be defined as the employee's emotional attachment to, or identification with, their firm (Meyer & Allen, 1997; Benligiray, & Sönmez, 2012) and their intention to continue working there (Rhoades et al., 2001; Allen et al., 2003; Bal et al., 2008; Robbins & Judge, 2013), and commitment is sometimes seen as an indicator of social well-being (Biétry & Creusier, 2015; Kara et al., 2015).

Contributions to the literature find negative relations between WFC and affective commitment (Allen et al., 2000; Liao, 2011) and between WFC and commitment to stay in the organization (Anderson et al., 2002; Boyar et al., 2003; Batt & Valcour, 2003; Premeaux et al., 2007; Carr et al., 2008).

The literature also points out that firms that favor WFP generate positive effects on organizational commitment (Allen, 2001; Kinnunen et al., 2005), because WFP encourages workers to develop more favorable attitudes and feelings about their work and their organization (Aryee et al., 2005; Wayne et al., 2006). Thus when firms alleviate the tension between work and family, worker turnover drops and worker retention increases (Glass & Estes, 1997; Drago et al., 2001; Bocaz, 2003; Foley et al., 2006; Estes et al., 2007; Swody & Powell, 2007). Likewise, empirical evidence exists that a lack of WFP generates low commitment and high absenteeism (Idrovo, 2006).

All this leads us to propose the following hypotheses:

H3: The greater the worker's perception that WFP exist in their firm, the greater their employee well-being.

H3a: The relation between the perception of the existence of WFP and employee well-being is stronger in women than in men.

H3b: The relation between the perception of the existence of WFP and employee well-being is stronger in workers with family responsibilities than in workers without family responsibilities.

H4: The greater the worker's perception that the WFP are accessible, the greater their employee well-being.

H4a: The relation between the perception of the accessibility of the WFP and employee well-being is stronger in women than in men.

H4b: The relation between the perception of the accessibility of the WFP and employee well-being is stronger in workers with family responsibilities than in workers without family responsibilities.

*Well-being and performance*

The literature offers evidence for a positive relation between employee well-being and job performance (Lyubomirsky et al., 2005). According to Wright (2014), one part of the literature defines well-being as satisfaction with life and finds a positive relation between well-being and job performance (Rode et al., 2005). More studies, however, examine well-being's impact on job performance using emotion-based well-being (Wright & Cropanzano, 2000; Wright et al., 2007). This emotion-based well-being exists when the worker does not suffer stress or emotional exhaustion and enjoys a positive affectivity (Wright, 2014). Positive affectivity means that the worker feels enthusiastic, active and alert, and predicts a readiness to solve conflicts, an optimistic mood, and a greater creativity and affiliative motivation.

Regarding the well-being and job performance relationship, where no significant differences between the groups are expected, the final hypothesis of this work is as follows:

> H5: The greater the worker's employee well-being, the greater their job performance.
>
> H5a: The relation between employee well-being and job performance is similar in men and women.
>
> H5b: The relation between employee well-being and job performance is similar in workers with family responsibilities and workers without family responsibilities.

Given the hypotheses proposed in this work, Figure 1 illustrates our model of the impact of WFP on job performance. This relation is mediated by employee well-being, which is measured by psychological indicators relevant in human resource management such as stress, motivation, satisfaction and commitment to the organization. Our model proposes positive relations between the existence and accessibility of WFP and employee well-being and job performance. It also proposes a direct, positive relation between well-being and job performance.

Figure 1. Theoretical model of WFP's impact on employee well-being and job performance

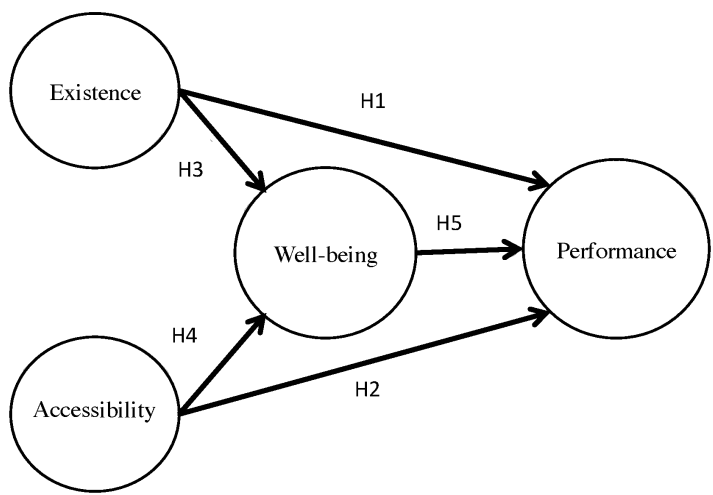

**Methodology**

*Sample and data collection*

The fieldwork for this research was carried out in the Spanish financial sector, which in 2014 comprised a total population of 192,265 employees. It was necessary to gather data in multiple organizations to increase the variance in work-family culture and WFP. Thus data were collected at thirteen Spanish organizations. We contacted the organizations for their participation, explaining our research purpose and offering a company-specific final report as an incentive to participate. All the information given to the companies was at the aggregate level and no respondent could be identified. To obtain the data we designed a questionnaire with closed questions. We carried out a pretest to improve its effectiveness and design, remove ambiguities and improve the measures. Employees were contacted by e-mail directly by the organizations. The type of survey carried out belongs to the category Internet-based self-administered survey. The survey was open from July to November 2014. A total of 1,565 questionnaires were thus collected, 54 of them were omitted due to missing or incomplete data. In the 1,511 effective responses, 42.4% of the respondents were female (641 women and 870 men); 75.9% of the participants had family responsibilities (1,147 respondents had family responsibilities and 364 did not); and the mean age was 43.7 (s.d. 8.9).

*Measures*

To measure the unobservable variables of the proposed model, existence and accessibility of WFP, well-being, and job performance, we have adapted validated scales from the literature (Families and Work Institute, 2012a, 2012b; Anderson et al., 2002; Boshoff & Mels, 2000; Warr, 1990). The evaluation of their psychometric properties (see Results section) confirms the reliability and validity of all of them.

All the constructs except job performance were measured using multiple indicators based on the scoring of a series of statements on 5-point Likert scales, with 1 indicating total disagreement and 5, total agreement. To measure job performance we used a single indicator on the same scale.

For all the latent variables we opted for a reflective measurement model because the indicators are manifestations of the unobserved, theoretical construct that they represent, they share a common theme, and they are exchangeable and highly correlated (Mackenzie et al., 2005).

Perceived existence of WFP was measured with five items from the Families and Work Institute (2012a, 2012b) Questionnaire (e.g. "I have the schedule flexibility I need at work to manage my personal and family responsibilities") ($\alpha$=0.783).

Perceived accessibility of WFP was measured with nine items from Anderson et al. (2002) and the Families and Work Institute (2012a, 2012b) Questionnaire (e.g. "If I used WFP, my manager wouldn't be supportive", reverse-scored; "If I used WFP, my co-workers wouldn't be supportive", reverse-scored; and "If I used WFP, there would be negative consequences for my job advancement", reverse-scored) ($\alpha$=0.895).

Employee well-being was measured with four items using scales validated by Anderson et al., (2002), Boshoff & Mels (2000), and Warr (1990) (e.g. "I often feel nervous and stressed", reverse-scored; or "The major satisfaction in my life comes from my work") ($\alpha$=0.963).

Perceived job performance was measured with a single item developed and validated by Boshoff & Mels (2000). The item was as follows: "My organization inspires the very best in me in the way of job performance" ($\alpha$: n.a.).

*Method*

To test our hypotheses we conducted a structural equation modeling. These models use both latent (unobservable) variables, which represent theoretical concepts, and data coming from measures (indicators or manifest variables), which are used as input for a statistical analysis to provide evidence about the relations between the latent variables (Williams et al., 2009). Following the generally accepted procedure for selecting the right model to estimate structural equation models (Hair et al., 2011; Hair et al., 2012b; Ringle et al., 2012), we opted for the PLS-SEM approach. To carry out the analysis we used the SmartPLS 3.0 software (Ringle et al., 2014), which uses the algorithm developed by Wold (1982) to estimate the parameters. This algorithm is essentially a sequence of regressions in terms of weighted vectors.

*Results*

Although the parameters of the measurement model and the structural model are estimated in a single step, we follow Chin's (2010) & Hair et al.'s (2011; 2012a,b,c; 2013; 2014) recommendations for the presentation of the results, first evaluating the measurement model and then evaluating the statistical significance of the model parameters. This ensures that we have valid and reliable measures before drawing conclusions about the relations between the constructs. Figure 2 reports the results of the PLS algorithm in SmartPLS 3.0.

Figure 2. Structural model adjustment. Factor loadings and path coefficients

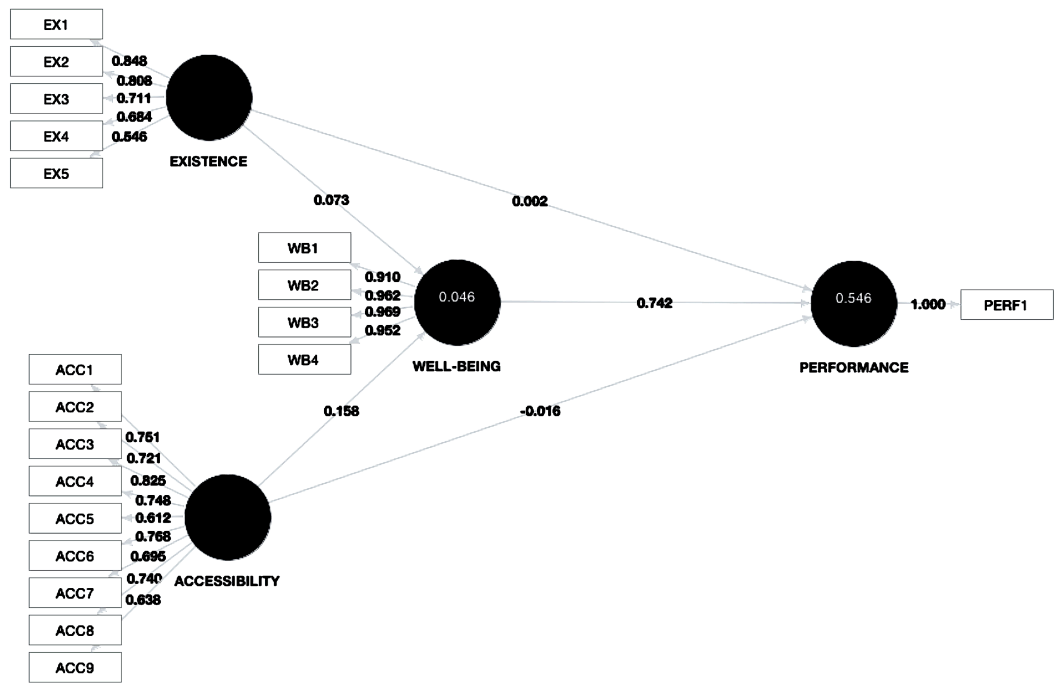

We use the mean substitution criterion to handle any missing data. This method has the advantage of maintaining the sample size and the mean of each variable though it could affect the variance and the correlation between the estimators.

*Evaluation of measurement model*

In this section we analyze whether the theoretical concepts are measured correctly via the observed variables or indicators. All the constructs in this model respond to a reflective model. First, we analyze the reliability, in other words, whether the indicators really measure what they are intended to measure; then we analyze the validity, or whether the measurement is done in a stable and consistent way.

In the reliability analysis we first evaluate the individual reliability of each item, examining the factor loadings (lambda) or simple correlations between the indicators and their respective latent variables. An indicator is accepted in the measurement model of a construct if it has a factor loading greater than or equal to 0.707. This means that the variance shared between the construct and its indicators is greater than the variance of the error (Carmines & Zeller, 1979). Some researchers argue that this empirical threshold ($\lambda \geq 0.707$) should not be so strict in the initial stages of scale development (Hair et al., 2014), and that indicators with loadings between 0.4 and 0.7 should only be eliminated from a scale if this leads to an increase in the average variance extracted (AVE) or composite reliability (CR) to over the suggested threshold value (AVE=0.5; CR=0.7). Thus weak indicators could be retained depending on their contribution to the content validity. Very weak indicators ($\lambda \leq 0.4$) should, however, always be eliminated.

In this research we retain all the indicators of the measurement model despite the fact that five of them do not meet the minimum required threshold ($\lambda \geq 0.707$). This decision is justified because the AVE of all the latent variables exceeds 0.5. This means that eliminating these indicators in an attempt to achieve the minimum AVE threshold would be pointless. Thus the indicators are retained because of their contribution to the content validity.

Second, we analyze the scale reliability. This process allows the researcher to ensure the internal consistency of all the indicators when measuring the concept, in other words, we evaluate how rigorously the indicators are measuring the same latent variable. To evaluate this aspect we use the Cronbach alpha coefficient and the composite reliability (Table 1). Nunnally (1978) suggests using 0.7 for a modest reliability and a stricter 0.8 for basic research. As Table 1 shows, all the constructs comfortably exceed these Cronbach alpha and CR values.

Table 1. Reliability and validity indicators

|        | Cronbach Alpha | Composite Reliability | AVE   | ACCESS | WELL  | PERF                | EXIST |
|--------|----------------|------------------------|-------|--------|-------|---------------------|-------|
| ACCESS | 0.895          | 0.908                  | 0.525 | **0.725** |       |                     |       |
| WELL   | 0.963          | 0.973                  | 0.900 | 0.207  | **0.949** |                 |       |
| PERF   | n.a.           | n.a.                   | n.a.  | 0.138  | 0.739 | Single-item construct |     |
| EXIST  | 0.783          | 0.846                  | 0.528 | 0.667  | 0.178 | 0.123               | **0.727** |

In turn, the validity analysis involves evaluating the convergent and discriminant validities. The first implies that a set of indicators represents a single, underlying construct, and this can be demonstrated by its unidimensionality (Henseler et al., 2009). This is evaluated by the AVE (Fornell & Larcker, 1981), which measures the share of the variance that a construct obtains from its indicators compared to that due to the measurement error. Fornell and Larcker (1981) recommend that the AVE exceeds 0.5, which would mean that more than 50% of the construct's variance is due to its indicators rather than the indicators in the rest of the constructs. As Table 1 shows, all the constructs have AVE values exceeding 0.5. Thus more than 50% of the variance of each construct is due to its indicators.

For the single-item construct PERF, the Cronbach Alpha, CR, and AVE are not appropriate measures since the indicator's outer loading is fixed at 1.00.

The discriminant validity evaluates the extent to which a given construct is different from other constructs. We follow Fornell and Larcker's (1981) criterion based on the idea that a construct should share more variance with its indicators than with the other constructs in a particular model. Thus a good way of measuring a construct's discriminant validity is by showing that its AVE is greater than the variance the construct shares with the other model constructs. Or in other words, we must show that all the correlations between the constructs are lower than the square root of the AVE.

As can be seen from the figures in bold in the table, for all the latent variables of the model the square root of the AVE is greater than all the correlations between the variables, which means that all the constructs are more strongly related to their own indicators than to those in the other constructs.

*Evaluation of structural model*

Having confirmed the goodness of fit of the measurement model, we analyze the structural model by evaluating the strength and significance of the relations between the different variables. In particular, this analysis involves evaluating the variance explained of the endogenous variables, measured by their $R^2$, their path coefficients or standardized regression weights (Beta), and their significance levels.

A measure of the predictive power of a model is the $R^2$ value of the dependent latent variables, which indicates the share of the construct's variance explained by the model. Chin (1998) recommends the following thresholds: from 0.67, "substantial"; from 0.33, "moderate"; and from 0.19, "weak". The $R^2$ value obtained in this model is weak for the variable Employee well-being ($R^2$=0.046), but moderate, close to substantial, for Job performance ($R^2$=0.542).

Second, the evaluation of the significance of the path coefficients is carried out using the bootstrapping resampling technique. The subsamples are created following the recommendations of the literature with 5,000 observations (Hair et al., 2014), extracted at random and with replacement of the original set of data. This is a sufficiently large number to ensure that the results are stable. Table 2 shows the results of the p-values. Apart from the above, and again following the recommendations of the literature we carry out an additional evaluation with a non-parametric test to calculate the confidence intervals.

Table 2. Results of structural model: Direct effects. Evaluation with t-values & confidence intervals

| Hypothesis | Expected effect | Path coefficient | t-value (bootstrap) | p-value | Confidence Interval Low | Confidence Interval Up | Support |
|---|---|---|---|---|---|---|---|
| H1: EXIST -> PERF The greater the worker's perception that WFP exist in their firm, the greater their job performance. | + | 0.002 | 0.075 | 0.940 | -0.047 | 0.052 | No |
| H2: ACCESS -> PERF The greater the worker's perception that the WFP are accessible, the greater their job performance. | + | -0.016 | 0.669 | 0.504 | -0.064 | 0.032 | No |
| H3: EXIST -> WELL The greater the worker's perception that WFP exist in their firm, the greater their employee well-being. | + | **0.073** | **2.167** | **0.030** | **0.008** | **0.139** | **Yes** |
| H4: ACCESS -> WELL The greater the worker's perception that the WFP are accessible, the greater their employee well-being. | + | **0.158** | **5.476** | **0.000** | **0.106** | **0.220** | **Yes** |
| H5: WELL -> PERF The greater the worker's employee well-being, the greater their job performance. | + | **0.742** | **42.039** | **0.000** | **0.707** | **0.775** | **Yes** |

For n = 5,000 subsamples: * $p < .05$; ** $p < .01$; ***$p < .001$ (based on 1-tailed t-Student distribution t(4999)). t(0.05; 4999) = 1.645; t(0.01; 4999) = 2.327; t(0.001; 4999) = 3.092. In bold, statistically significant values

In addition, to evaluate the possible indirect and total effects between the latent variables we repeat the previous procedure for these relations (see Table 3).

Table 3. Indirect and total effects: Evaluation with t-values & confidence intervals

| | | Path coefficient | Standard Error (STERR) | T Statistic (|O/STERR|) | p-value | Confidence Interval Low | Confidence Interval Up |
|---|---|---|---|---|---|---|---|
| Indirect effects | EXIST -> PERF | **0.054** | **0.025** | **2.156** | **0.031** | **0.006** | **0.104** |
| | ACCESS -> PERF | **0.117** | **0.022** | **5.399** | **0.000** | **0.079** | **0.164** |
| Total effects | ACCESS -> WELL | **0.158** | **0.029** | **5.476** | **0.000** | **0.106** | **0.220** |
| | ACCESS -> PERF | **0.101** | **0.031** | **3.254** | **0.001** | **0.042** | **0.166** |
| | WELL -> PERF | **0.742** | **0.018** | **42.039** | **0.000** | **0.707** | **0.775** |
| | EXIST -> WELL | **0.073** | **0.034** | **2.167** | **0.030** | **0.008** | **0.139** |
| | EXIST -> PERF | 0.056 | 0.034 | 1.648 | 0.099 | -0.010 | 0.121 |

For n = 5,000 subsamples: * $p < .05$; ** $p < .01$; ***$p < .001$ (based on 1-tailed t-Student distribution t(4999)).

t(0.05; 4999) = 1.645; t(0.01; 4999) = 2.327; t(0.001; 4999) = 3.092. In bold, statistically significant values.

The results of the analysis do not provide support for Hypothesis 1 (ß=0.002; p-value=0.940) or Hypothesis 2 (ß=-0.016; p-value=-0.504), as Table 2 shows. This implies that the mere existence and accessibility of the WFP do not generate an improvement in performance. Nevertheless, these two variables do have an indirect effect on performance via employee well-being (ß=0.054; p-value=0.031; and ß=0.117; p-value=0.000, respectively), as Table 3 shows. Moreover, adding the direct and indirect effects we find that accessibility does have a total effect on performance (ß=0.101; p-value=0.001), though existence does not (ß=0.056; p-value=0.099).

As Table 2 shows, the results do provide support for Hypothesis 3 (ß=0.073; p-value=0.030) and Hypothesis 4 (ß=0.158; p-value=0.000). This implies that existence and accessibility have very weak but statistically significant effects on employee well-being.

Finally, the results also provide support for Hypothesis 5 (ß=0.742; p-value=0.000): employee well-being generates improvements in performance. In fact, of the three variables expected to have a direct effect on performance in our proposed model this is the only one for which we have found empirical evidence.

*Analysis of moderating role of gender and family responsibilities*

As mentioned in the theoretical framework section and given the nature of the phenomenon of interest, we suspected that there would be heterogeneity in the data due to gender and the existence of family responsibilities.

To analyze the possible moderating role of gender and family responsibilities we carry out two multigroup analyses. Multigroup analysis (MGA) is a method to test for significant differences between subsamples. To analyze whether significant differences exist in the factor loadings and path coefficients between the two groups we use Henseler's (2007) test. This is a non-parametric significance test that compares the differences in the specific results of each group on the basis of the PLS-SEM bootstrapping results. A difference is significant at the 5% probability of error if the p-value is less than 0.05 or greater than 0.95 for a particular difference in values between the groups.

Before carrying out the test for significant differences in the parameters we evaluate the metric invariance of the measurement model in each subsample in order to validate the goodness of fit of the measurement model for each subsample.

The first MGA is carried out with 641 women and 870 men and the second with 1,147 respondents with family responsibilities and 364 respondents without family responsibilities. We do not report these results for space reasons, but all the indicators analyzed—except for one in the comparison between men and women—show that no significant differences exist in the measurement model, so we can accept the hypothesis of metric invariance.

We then run Henseler's test to evaluate differences in the path coefficients of the structural model in both subsamples (Table 4 and Table 5). A resampling of 5,000 cases is carried out and the results form the basis for the test for significant differences between the groups. The results of each resampling for the second group are compared to the results for the first group. The number of positive, significant differences divided by the total number of comparisons indicates the probability that the parameter from the second group is higher than the parameter from the first group (Afonso et al., 2012).

As Table 4 and Table 5 show, all the p-values are greater than 0.05, which means that no statistically significant differences exist in the relations analyzed between men and women or between respondents with or without family responsibilities.

Table 4. Henseler's test for multi-group analysis: evaluation of significant differences in model due to gender

| Hypotheses: Women vs. Men | Path Coefficients-diff | p-value | Support |
| --- | --- | --- | --- |
| H1a: EXIST -> PERF<br>The relation between the perception of the existence of WFP and job performance is stronger in women than in men. | 0.010 | 0.426 | No |
| H2a: ACCESS -> PERF<br>The relation between the perception of the accessibility of the WFP and job performance is stronger in women than in men. | 0.002 | 0.510 | No |
| H3a: EXIST -> WELL<br>The relation between the perception of the existence of WFP and employee well-being is stronger in women than in men. | 0.016 | 0.586 | No |
| H4a: ACCESS -> WELL<br>The relation between the perception of the accessibility of the WFP and employee well-being is stronger in women than in men. | 0.029 | 0.618 | No |
| H5a: WELL -> PERF<br>The relation between employee well-being and job performance is similar in men and women. | 0.009 | 0.604 | No |

Table 5. Henseler's test for multi-group analysis: evaluation of significant differences in model due to family responsibilities

| Hypotheses: With responsibilities vs. Without responsibilities | Path Coefficients-diff | p-value | Support |
| --- | --- | --- | --- |
| H1b: EXIST -> PERF<br>The relation between the perception of the existence of WFP and job performance is stronger in workers with family responsibilities than in workers without family responsibilities. | 0.007 | 0.452 | No |
| H2b: ACCESS -> PERF<br>The relation between the perception of the accessibility of the WFP and job performance is stronger in workers with family responsibilities than in workers without family responsibilities. | 0.048 | 0.173 | No |
| H3b: EXIST -> WELL<br>The relation between the perception of the existence of WFP and employee well-being is stronger in workers with family responsibilities than in workers without family responsibilities. | 0.030 | 0.352 | No |
| H4b: ACCESS -> WELL<br>The relation between the perception of the accessibility of the WFP and employee well-being is stronger in workers with family responsibilities than in workers without family responsibilities. | 0.049 | 0.774 | No |
| H5b: WELL -> PERF<br>The relation between employee well-being and job performance is similar in workers with family responsibilities and workers without family responsibilities. | 0.110 | 0.999 | No |

**Discussion**

These findings have some theoretical implications. The contribution of this work to literature is twofold. On the one hand, it deals with the unpublished role of the variable WB as a mediator of the relationship between WFP and job performance. On the other hand, we have considered two different dimensions of WFP to measure their impact on job performance. The first dimension is the knowledge that WFP exist (Yeandle et al., 2002, Haar & Spell, 2004, Mumford, 2005). The second dimension is the perception that these WFP are accessible without negative consequences for workers (Gray & Tudball, 2002, Bond, 2004). Although some studies differentiate these two dimensions of WFP (Gray & Tudball, 2002, Yeandle et al., 2002, Haar & Spell, 2004, Bond, 2004, Mumford, 2005), none of them empirically analyzes this difference and their impact on job performance.

In this research we have shown that employee well-being has a mediating effect in the relation between work-family policies (WFP) and job performance. This is true for the two dimensions of WFP considered here: existence and accessibility. Our concept of well-being is basically founded on the emotion-based well-being approach (Wright, 2014). A large number of studies have shown that WFP help to reduce stress, negative emotions and affective states (dissatisfaction, lack of motivation, or lack of commitment) and unfair situations related to work-family conflict (WFC). Thus WFP contribute to improving employee well-being and consequently job performance (Beauregard & Henry, 2009; Yasbek, 2004; Matta et al., 2014; Penney & Spector, 2005; Spector et al., 2006).

      These results are coherent with Allen, French, Dumani, and Shockley's (2015) meta-analysis. These authors state that the role of the organization is stronger in avoiding work interference with the family (WIF) than in avoiding family interference with the work (FIW), with the latter being influenced more by the environment. WFP implemented by firms may positively impact on WIF conflicts, which explains why they increase the workers' well-being. However, the increase in FIW conflicts depends more on environmental factors like the existence of a collectivistic (versus

individualistic) culture or an economic gender gap (Allen et al., 2015). Thus, firms' efforts to reduce conflicts with WFP may not impact directly on performance, as firms do not have much control over FIW, but may do so indirectly mediated by workers' well-being.

On the other hand, we have failed to find empirical evidence here that gender and family responsibilities have a moderating role on the relations in our model, contrary to what we initially expected. These results contrast with the results of previous studies such as Kossek et al. (2001), who find that family responsibilities affect well-being and performance, and Harris (2004), who shows that women suffer more stress and are less satisfied with their work—both components of well-being—as a result of WFC. In fact, various studies find a negative relation between WFC and satisfaction among women (Bruck et al., 2002; Kossek & Ozeki, 1998). However, our results are coherent with other studies failing to find significant differences in the impact of gender on conflict and well-being (Akintayo, 2010; Hughes & Bozionelos, 2007). In our particular study, this could be due to the organizational culture of the sector we study —the financial sector— or to the cultural characteristics of the country concerned —Spain.

Most of the studies on WFC and WFP have been carried out in Anglo-Saxon countries (Idrovo et al., 2012). But the culture and specific values of each country have an important impact on WFP existence and accessibility and on their impact (Karatepe & Uludag, 2008). WFP may differ among countries due to historical, political, and attitudinal reasons (Notten, Grunow, & Verbakel, 2017). For instance, in Muslim cultures women's participation in the labor market is low (Cohen, Granot & Yishai, 2007). Similarly, most Latin American countries have a low rate of female labor participation (Jiménez & Gómez, 2015; Idrovo, 2006). In these contexts, there is a poor culture of WFP (Idrovo, 2006). On the other hand, the USA lags behind in terms of WFP, despite its higher rate of female participation in the labor market (Hammer & Zimmerman, 2011; Ray, Gornick & Schmitt, 2009; Heymann, Earle & Hayes, 2007). Various US studies point to the lack of support for working families and defend the adoption of WFP in the European style (Williams, 2010). Nevertheless, the development of WFP in Europe is heterogeneous (Adame-Sánchez & Miquel-Romero, 2012).

WFP have been a priority objective of the social policies of some European countries for decades, even before they became a key element of European Union policies. However, in other European countries, especially the Mediterranean ones, these policies have been imposed by European regulations. Thus, there is a lack of WFP tradition in Spain and there are cultural and institutional differences with respect to other European countries (Moreno, 2011).

Following previous research, countries can be classified as liberal welfare state, conservative, social-democratic or postcommunist welfare (Notten et al., 2017). Spain can be included in the conservative group. The WFP in the conservative welfare regimes usually include very long family leave and provide a job guarantee for the parent on leave (Notten et al., 2017). In the case of Spain, parents can share all or part of many of these benefits (e.g. maternity and paternity leave), which may explain the lack of differences in our model due to gender. In addition, a number of WFP in Spain are also extended to workers without family, which may explain the lack of differences in our model due to family responsibilities. Our findings are consistent with those of Glass, Simon, & Andersson (2016), for whom in countries like Spain WFP are associated with smaller disparities in well-being between parents and non-parents. Moreover, the WFP that increase parental well-being do not reduce the well-being of nonparents.

Another factor that may also influence the lack of differences in our model due to gender and family responsibilities is the collectivistic nature of Spain. According to Allen et al. (2015), Asian and Latin American societies are considered collectivistic while Anglo Saxon societies are more individualistic. Individualists tend to focus on the self and on personal achievement. For historical reasons, Spain has important cultural similarities with the countries of Latin America. The collectivistic culture present in Spain reduces differences due to gender or family responsibilities. According to Allen et al. (2015), WFC is less of a concern in collectivistic countries than in individualistic countries. In the former, individuals seek help from family members (Allen et al., 2015), and employers support their employees more in exchange for their loyalty (Powell et al., 2009).

*Implications for management*

From the manager's perspective, the mediating role of well-being in the relation between WFP and job performance implies on the one hand that management can improve employee well-being by offering WFP, publicizing their existence and making sure that they are accessible without reprisals of any kind; and on the other, that this well-being can ultimately improve performance. In this respect, the results obtained in this study show that well-being explains an important part of performance. Thus, introducing policies that improve employee well-being could lead to important improvements in job performance and ultimately firm performance. But for this to happen human resource managers must ensure that the WFP exist, that the workers know about them, and that they are accessible.

Nevertheless, well-being is only partially explained by WFP. Managers should also bear in mind other variables that can also affect well-being. For example stress due to overwork could reduce well-being (Harris, 2004; Martha, 2010; Viitala, Tanskanen & Säntti, 2015; Le Fevre, Boxall & Macky, 2015), but WFP cannot reduce this stress because it has nothing to do with WFC. Similarly, remuneration (Vosloo, Fouché & Barnard, 2014) and working conditions (Guzi & de Pedraza, 2015; Robone, Jones & Rice, 2011) among other factors have important effects on well-being .

Another important question to consider is that firms that apply WFP can achieve greater performance improvements in countries with a deficient legislation in this area because their workers can experience a greater relative well-being.

**Conclusions**

This work has analyzed the effects of work-family policies on well-being and job performance. For this purpose, we have considered two dimensions of WFP: (1) whether these policies exist and are known about by the employees; and (2) whether they are in practice accessible to the workers without reprisals. We have also looked for evidence for a moderating role of gender and family responsibilities. The results of this analysis show that the mere existence and accessibility of WFP do not generate improvements in performance. On the other hand, these two variables do have an indirect effect on performance via employee well-being. Moreover, and again according to the results, neither gender nor family responsibilities has a moderating role on the above relations, contrary to what we initially expected.

*Limitations*

One of the limitations of this study is the difficulty in measuring well-being (Wright, 2014). Although we can say that the indicators we use are components of well-being, they do not explain all the possible dimensions of this concept. Thus the concept of well-being used in this work is more a contribution to it than an explanation.

The limitations of the fieldwork include the difficulty in encouraging employees from the sector of interest to respond to our questionnaire. Finance workers are usually very busy, they have limited free time and their sector was in the midst of a crisis. In addition, the content of the questionnaire could have encouraged a disproportionate number of the most disgruntled workers to respond to the questionnaire and at the same time discouraged workers fearing possible reprisals from their firm. Moreover, the same respondents answered all the questionnaire items. Therefore, common method variance (CMV) might exist because the data used were from the same source. To examine CMV, Harman's one-factor test (Anderson & Bateman, 1997) could be employed in future research.

Another limitation of the results of this study is that the sample comes from one specific sector—the finance sector—and one single country—Spain. The normative specificities (with regard to the legislation on WFP) and cultural idiosyncrasies of this sample mean that any generalization to other sectors and countries must be carried out with caution.

Finally, the cross-sectional nature of the study is another limitation. As this study is cross-sectional it represents a first approach that needs to be complemented by a longitudinal study.

*Future lines of research*

It would be interesting to replicate this analysis in other sectors and countries. Regarding the international comparison, the collectivistic versus individualistic nature of different countries, as well as the existence or absence of gender egalitarianism, should be taken into account in any analysis of differences in the WFP-wellbeing and wellbeing-performance relationships due to gender or family responsibilities. Studying the influence on the model of economics variables such as GDP, unemployment rate or economic gender gap (Allen et al., 2015) may also be valuable. Such a comparative analysis could have important implications for firm managers and for the establishment of national policies in favor of WFP.

As we have mentioned, WFP can explain well-being only partially. Researchers should consider other variables affecting well-being as control variables in the model. These could include demographic, socio-economic, sectorial, geographic, legislative and cultural variables.

Finally, as mentioned in the Limitations section, this research should be accompanied by a longitudinal study, since the effects of WFP on employees' well-being and performance may evolve over time and so the relations considered in the model may change.

**Acknowledgments**